\documentclass[11pt]{article}

\usepackage[margin=1in]{geometry}
\usepackage{amsmath,amssymb,amsthm,mathtools}
\usepackage{enumitem}
\usepackage{setspace}
\usepackage{color}
\usepackage[round]{natbib}
\usepackage[hidelinks]{hyperref}
\usepackage{cleveref}

\newcommand{\E}{\mathbb E}
\newcommand{\Prb}{\mathbb P}
\newcommand{\R}{\mathbb R}
\newcommand{\CS}{\mathrm{CS}}
\newcommand{\OPT}{\mathrm{OPT}}
\newcommand{\SEP}{\mathrm{SEP}}

\newcommand{\MB}{\mathrm{MB}}
\newcommand{\BNE}{\mathrm{BNE}}

\newcommand{\barB}{\bar B}
\newcommand{\barh}{\bar h}

\newtheorem{theorem}{Theorem}
\newtheorem{lemma}{Lemma}

\theoremstyle{definition}
\newtheorem{definition}{Definition}
\newtheorem{example}{Example}
\theoremstyle{remark}
\newtheorem{remark}{Remark}

\crefname{theorem}{Theorem}{Theorems}
\crefname{lemma}{Lemma}{Lemmas}
\crefname{corollary}{Corollary}{Corollaries}
\crefname{definition}{Definition}{Definitions}
\crefname{example}{Example}{Examples}
\crefname{remark}{Remark}{Remarks}
\crefname{section}{Section}{Sections}
\crefname{equation}{}{}

\title{Artificial Currency and Separate Selling:\\
A Constant Approximation for Consumer Surplus\thanks{This paper is developed with the assistance of ChatGPT. Yang Cai was supported by the NSF Awards CCF-1942583 (CAREER) and CCF-2342642. Yingkai Li thanks NUS Start-up Grant for financial support. Jinzhao Wu was supported by the NSF Awards CCF-1942583 (CAREER), CCF-2342642, and a CADMY Research Fellowship. }}

\author{Yang Cai\thanks{Department of Computer Science, Yale University. Email: \url{yang.cai@yale.edu}} \\ 
\and Yingkai Li\thanks{Department of Economics, National University of Singapore. Email: \url{yk.li@nus.edu.sg}}
\and Jinzhao Wu\thanks{Department of Computer Science, Yale University. Email: \url{jinzhao.wu@yale.edu}} }

\date{}

\begin{document}
\maketitle

\begin{abstract}
We study the allocation of indivisible items to buyers with submodular values
generated by independent item signals. The objective is consumer surplus net
of burnt payments. We compare two simple mechanisms. \emph{Separate selling}
applies money burning item by item: each item is sold in its own auction
with burnt payments. \emph{Artificial-currency mechanisms} give
buyers personalized budgets of a currency worthless outside the
mechanism, which they use to bid across items. The better of our separate-selling
and artificial-currency mechanisms provides a \(3.80\)-approximation to
optimal consumer surplus. For additive buyers, the factor improves to \(3\).
The artificial-currency guarantee holds at \emph{every}
Bayes--Nash equilibrium, without equilibrium selection. Either approach alone
may capture only a vanishing share of the optimum. The proof decomposes
benchmark surplus into a stable part captured by artificial currency and a
variation bounded using separate selling.
\end{abstract}


\section{Introduction}
\label{sec:intro}

Some of the most consequential allocation problems are run without money.
Business schools assign seats in oversubscribed courses by giving students a
fixed allowance of points to bid \citep{SU10,Budish11}; Feeding America
allocates donated food by auctioning truckloads to food banks for an
artificial currency of ``shares'' \citep{Prendergast22}; organizations ration
computing time, instruments, and staff across their own teams; public
programs place applicants into scarce slots. Charging is ruled out --- by
law, by fairness, or by the designer's own objective --- yet values are
private and the goods are scarce, so these systems still need what prices
provide: a way to learn who wants what, and how much. And the goods are many.
A student assembles a schedule from many seats, a food bank stocks a larder
from many loads, a manager divides a cluster across many projects; what a
participant cares about is a bundle, and what the designer divides is many
heterogeneous goods at once.

Practice answers with two devices. One is the ordeal: make people
\emph{prove} that they want the good by wasting something --- queue for it,
fill out forms for it, sit through delays for it. Ordeals are payments in
everything but bookkeeping: they screen exactly as prices do, but what they
collect is destroyed rather than transferred. \citet{HR08} call this
\emph{money burning}, and it presents the designer with a real trade-off:
burnt payments extract information, and every unit extracted is a unit of
surplus destroyed. Her natural objective is therefore \emph{consumer
surplus} --- the value her allocation delivers net of the utility her
screening destroys. The other device is the one the course bidders and the
food banks use: print an allowance of play money, worthless outside the
mechanism, and let each participant spend it as she sees fit. Nothing is
burnt there, because nothing real is ever paid.
But the allowance is fixed before a participant learns her values: she can
direct her spending toward the items she values most, yet her budget does
not grow when her values are unusually high.

Each device thus has its own limitation. Money
burning can help identify buyers with high values, but does so by imposing
real costs: time spent waiting in a queue, for example, is lost. Artificial
currency avoids these costs, but it reveals only the relative strength of
a buyer's preferences across goods, not their overall level. How much of
the best achievable consumer surplus can the two devices deliver?

Classical results characterize optimal money burning in single-parameter
environments \citep{HR08,Condorelli12,CK13}, whereas the applications above
involve multidimensional preferences over heterogeneous goods.
We study the multidimensional problem with \(n\) buyers who have submodular values for \(m\)
indivisible items.
Each buyer privately observes one signal per item. Her value for a bundle
depends on the signals of the items it contains and is monotone submodular:
increasing one signal raises value by a weakly smaller amount when the
other signals are higher. This class includes additive and unit-demand values. All buyer-item signals are drawn independently from
distributions with finite support. The distributions may differ across buyers
and items, and need not be regular. Our benchmark
\(\OPT\) is the maximum of expected consumer surplus over all Bayesian
incentive-compatible mechanisms with interim individual rationality and
nonnegative payments at every report profile; all payments are burnt.

We compare this benchmark with two simple mechanisms, one for each device.
\emph{Separate selling} applies money burning to each item in its own
auction. With additive values, a buyer keeps every item she wins. With
submodular values, we use \emph{one-sale separate selling}: the auctions are
chosen jointly and each buyer takes only her best offer, which preserves
truthful reporting because she receives at most one item. Our
artificial-currency mechanism is the \emph{proportional credit mechanism}:
each buyer receives a personalized budget of credits, spreads it across
items after observing her signals, and wins each item with probability
proportional to the credits placed on it.

Our main result is that the two failures do not overlap: on every instance one
device or the other performs, and together they reach a constant fraction of
the unrestricted optimum.

\medskip
\noindent\textbf{Theorem} (informal)\textbf{.}
\emph{For every product prior with finite support, the better of one-sale
separate selling and a proportional credit mechanism provides a
\(3.80\)-approximation to optimal consumer surplus.
With additive values and unrestricted separate selling, the factor can be
made arbitrarily close to \(3\). The credit mechanism's guarantee holds at every
Bayes--Nash equilibrium, without equilibrium selection.}
\medskip

We evaluate separate selling under truthful reporting, which is a
dominant strategy.\footnote{An arbitrarily small random perturbation makes truthful reporting
strictly dominant, yielding a unique equilibrium with arbitrarily small
loss in expected consumer surplus.} We require the artificial-currency guarantee to hold at every
equilibrium, as in the price-of-anarchy literature
\citep{Roughgarden15,ST13}.\footnote{By the revelation principle, each
Bayes--Nash equilibrium can be reproduced by a Bayesian incentive-compatible
direct mechanism with the same expected consumer surplus, preserving our
guarantee \citep{Myerson81}. We present the indirect format because its bidding
rules are easy to understand and implement in practice.}

Both approaches are needed for the guarantee.
With a single item and no payments, all strictly positive values rank
allocation lotteries identically, so artificial currency cannot screen across value levels.
When high values are rare,
every artificial-currency mechanism therefore captures only a vanishing fraction
of \(\OPT\) at any
equilibrium it admits (\cref{ex:credit-fails}). Conversely, when values are
i.i.d.\ with truncated geometric tails, separate selling captures a vanishing fraction of
\(\OPT\) (\cref{ex:sep-fails}).

\Cref{sec:overview} outlines the argument and \cref{sec:lit} discusses
related work. \Cref{sec:model} defines the model and the two mechanisms.
\Cref{sec:main} states the approximation guarantees, gives examples
showing why both mechanisms are needed, and reduces the guarantees to two
lemmas. \Cref{sec:deviation,sec:variation}
prove the two lemmas, and \cref{sec:conclusion} discusses the implications
and possible extensions. The appendices complete the submodular variation argument and prove
equilibrium existence.

\subsection{Overview of the argument}
\label{sec:overview}

\paragraph{Stable part and variation.}
The proof begins by decomposing the surplus of an arbitrary benchmark
mechanism. For buyer \(i\), let \(u_i(s_i)\) be her interim utility, and
draw an independent copy \(s_i'\) of her type \(s_i\). Define the
\emph{stable part} as
\(K_i=\E\min\{u_i(s_i),u_i(s_i')\}\) and the \emph{variation} as
\(D_i=\tfrac12\E|u_i(s_i)-u_i(s_i')|\). Together they account for her expected
utility: \(K_i+D_i=\E[u_i(s_i)]\). The variation is half the Gini mean
difference of her interim utility, so the stable part is her expected
utility less a measure of its dispersion across types.
For example, a buyer whose utility is \(N\) with probability \(1/N\) and
zero otherwise has expected utility \(1\) but stable part \(1/N\): almost
all of her surplus is variation.
For the bounds, we use an upper bound, the \emph{stable value}
\(\widehat K_i=(\E\sqrt{w_i(s_i)})^2\ge K_i\), where \(w_i(s_i)\)
is her interim gross value under the benchmark. Writing \(\widehat K\) and
\(D\) for the sums over buyers, the benchmark's consumer surplus is at most
\(\widehat K+D\). Two lemmas then
establish the result: our artificial-currency mechanisms capture stable
value, while separate selling controls the variation.

\paragraph{Credits capture stable value.}
A fixed allowance helps
provide the surplus a buyer receives consistently across types.
In the proportional credit mechanism, we choose \(B_i\propto\widehat K_i\), giving each buyer a
budget proportional to her stable value. These budgets depend
on the benchmark and prior, but not on buyers' realized values.
Each buyer also receives a small seed credit on every item; the seeds have
total mass \(\eta\) relative to the budgets. They serve purely to ensure
that an equilibrium exists, and \(\eta\) can be made arbitrarily small.

We interpret an equilibrium of this mechanism as a market. The average behavior
of buyer \(i\)'s opponents determines an implicit price \(q_{ij}\) for each item,
measured in artificial currency. Independence across buyers makes these
prices independent of her own type. For any nonempty target bundle \(A\) drawn from the benchmark, she can
bid to secure each item in the bundle with probability at least
\(B_i/(B_i+\sum_{j\in A}q_{ij})\). By submodularity, this also secures
at least the same fraction of the bundle's value in expectation.
Feasibility bounds the expected total implicit cost of these target
bundles across buyers by the total budget plus the seeds.
This deviation argument yields a smoothness-style bound on a weighted sum
of ratios between benchmark gross values and equilibrium utilities, and
applies at every equilibrium. Cauchy--Schwarz then converts the ratio bound
into a surplus guarantee: at every equilibrium, the consumer surplus
\(\CS\) of the credit mechanism satisfies
\((2+\eta)\,\CS\ge\widehat K\). The square root in \(\widehat K_i\) reflects the
limitation of a fixed budget: because one budget must serve every type, it captures the reliable component of a
buyer's benchmark gross value rather than its full mean.

\paragraph{Separate selling covers the variation.}
Money
burning can instead target unusually high values, because buyers with these
values are more willing to incur screening costs.
For additive values, we show that \(D\le\SEP\), where \(\SEP\) is the
consumer surplus of optimal separate selling. Fix a buyer and replace her signal
coordinates one at a time with independent draws. Incentive compatibility makes her interim
utility coordinatewise nondecreasing. For a monotone function of
one coordinate, half the expected absolute difference between two
independent draws is bounded by the expected utility difference from
setting that coordinate to zero. For each item, the expected coordinate utility differences,
summed across buyers, give the consumer surplus of a
feasible single-item mechanism obtained by integrating out the other
coordinates of the benchmark. Summing over items therefore bounds the
total variation by separate selling.

With submodular values, two things change. First, a buyer takes only her best offer, so
one-sale separate selling, whose optimal surplus we denote by \(\SEP_1\),
covers a maximum of coordinate utility differences rather than their sum.
Second, the sum can be much larger than the maximum. This happens when many
items each matter a little. But then the buyer's utility is concentrated:
if \(k\) coordinates matter equally, her variation is at most of order \(\sqrt k\)
times one coordinate's utility difference, rather than \(k\) times, and it
can be charged to her stable value. A product-space
inequality makes this precise and gives
\(D\le0.382\,\widehat K+1.03\,\SEP_1\).

\paragraph{Combining the steps.}
For additive values, the three steps give
\[
 \OPT\le\widehat K+D
 \le(2+\eta)\,\CS+\SEP
 \le(3+\eta)\max\{\CS,\SEP\}.
\]
The factor \(3\) is thus \(2\) for stable value plus \(1\) for the
variation; the additional \(\eta\) is the cost of the seeds and vanishes
as \(\eta\) tends to zero. For submodular values, the same chain gives
\(\OPT\le1.382\,\widehat K+1.03\,\SEP_1\) and the factor
\(1.382\cdot2+1.03<3.80\) once \(\eta\) is small. The deviation argument uses
independence across buyers; the variation argument additionally uses
independence across each buyer's item signals.

\subsection{Related literature}
\label{sec:lit}

\citet{HR08} develop a framework for maximizing consumer surplus net of burnt
payments and characterize optimal mechanisms
in single-parameter settings. \citet{Condorelli12} studies allocation through
costly signals and identifies when screening is preferable to allocation by
lottery. \citet{CK13} consider settings in which wasteful signaling is the
only available instrument and give conditions under which a lottery is
optimal. The single-item results of this literature underpin our separate-selling
mechanisms.
More recently, \citet{ESS25} study consumer-surplus maximization in
multi-parameter environments, including unit-demand buyers (whose value for a
bundle is the highest of their values for its items) with heterogeneous
items, and obtain tight approximation guarantees relative to social welfare.
\citet{GL25} study unit-demand buyers with i.i.d.\ values
and obtain tight welfare-approximation guarantees. 

In concurrent work,
\citet{GLT26} obtain the first constant-factor approximation to optimal consumer surplus with burnt payments for multiple matroid-constrained additive buyers with independent item values. 
They use an ex-ante relaxation to study each buyer separately under bounds on her expected item allocations, then combine the resulting mechanisms through contention resolution. 
They also give polynomial-time constructions under oracle-access assumptions, including access to expected free-lottery utilities for matroid-constrained buyers.
\citet{GMP26} show that exact consumer-surplus maximization under ex-ante allocation constraints can require an infinite menu even for one unit-demand or additive buyer with two independent item values.
We study monotone submodular values generated by independent
item signals, a class that includes additive, unit-demand, and
matroid-constrained additive values.
Our guarantee combines separate selling with our artificial-currency
mechanisms.

Pseudo-markets with artificial budgets date
to \citet{HZ79}, whose price-taking equilibria allocate probability shares
efficiently. \citet{Budish11} studies approximate competitive equilibrium from
equal incomes for indivisible bundles.
\citet{BCKO17} develop and implement Course Match, adapting this approach to
large-scale course allocation.
\citet{SU10} analyze the bidding-points systems used by business schools, and
\citet{Prendergast22} documents Feeding America's allocation of donated food
at national scale using a currency of shares.
\citet{AS16} study nonmonetary allocation of heterogeneous services with
private multidimensional utilities and known priors, optimizing a large-market
approximation and translating it into a finite-market mechanism. Our focus
is a uniform approximation to optimal consumer surplus with burnt payments
for buyers with monotone submodular values.

A related literature studies repeated nonmonetary allocation.
\citet{GBI21} use artificial currency to convert monetary mechanisms into
repeated nonmonetary mechanisms, with performance losses and gains from
misreporting that vanish as the horizon grows. Their results motivate our use of an artificial-currency mechanism for the stable part of benchmark surplus: a budget disciplines a buyer's total spending, over rounds in their setting and over items in ours. \citet{BGS19} use promises of future allocations to induce truthful reports
and approach first-best welfare as agents become patient. \citet{BJ26}
characterize distribution-dependent convergence rates to first best in finite
and discounted infinite horizons using promised utilities. These approaches
use repeated interaction to provide incentives. Our budgets instead link
simultaneous choices across heterogeneous items, and our guarantee holds at
every Bayes--Nash equilibrium in a one-shot market. Sharing a budget across items also relates to the linked decisions
of \citet{JS07} and the intertemporal budgets in the storable-votes model of
\citet{Casella05}.

\citet{Kelly97} studies
proportional sharing of a divisible resource. In the complete-information
setting with concave utilities, \citet{JT04} show that its worst-case
social-welfare loss is \(25\%\). \citet{BGM22} analyze trading post, in which agents distribute fixed budgets
across divisible goods and receive shares proportional to their bids. They
obtain a factor-two approximation to optimal weighted Nash social welfare
for concave utilities at every pure Nash equilibrium of the complete-information
game. Their extension to mixed equilibria imposes additional conditions on
satiation and competition \citep[Theorem~11]{BGM22}. Our proportional credit mechanism is a
seeded Bayesian version of this rule, interpreting shares as allocation
probabilities. Their proof uses a scaled-allocation deviation and a
budget-weighted utility-ratio bound. We extend this argument to Bayesian
bundle targets and choose budgets from the benchmark's stable values.
Our central inequality uses a
smoothness argument in the sense of \citet{Roughgarden15} and
\citet{ST13}: a single deviation bounds the benchmark against every
equilibrium. This is why our guarantee does not depend on equilibrium
selection.

Our comparison of two simple formats with an unrestricted optimum is related
to the literature on simple mechanisms for revenue maximization
\citep{HartNisan17}. For a single additive buyer with independent item values,
\citet{BILW20} show that the better of separate selling and grand bundling
achieves a constant-factor approximation to optimal revenue. This guarantee
has been sharpened in the additive setting \citep{MSL21,CaiEtAl26} and
extended to a single subadditive buyer with independent items \citep{RW18}.
With multiple additive buyers and independent item values, the better of
separate selling and a Vickrey--Clarke--Groves auction with buyer-specific
entry fees that may depend on the other buyers' reports also gives a
constant-factor approximation \citep{Yao15,CDW21}.
\citet{CDW21} develop a unified duality framework for these guarantees, which
\citet{CZ17} extend to broader valuation classes.
Our guarantee has the same form, with consumer surplus net of burnt
payments in place of revenue: separate selling plays the same role, and
artificial currency takes the place of bundling.

\section{Model}
\label{sec:model}

\subsection{Values and information}

A designer allocates \(m\ge1\) indivisible items among \(n\ge1\) buyers.
Buyers privately observe signals that determine their values, and the designer seeks to maximize
the value they receive net of any payments, which are burnt. We write
\([n]=\{1,\ldots,n\}\) and \([m]=\{1,\ldots,m\}\), using \(i\) for buyers
and \(j\) for items.

Buyer \(i\)'s type is a vector of item signals
\(s_i=(s_{ij})_{j\in[m]}\). Each signal takes values in a finite set
\(\mathcal S_{ij}\subseteq\R_+\). The signals \(s_{ij}\sim G_{ij}\) are
independent across buyers and items, and their distributions are commonly
known. Each \(\mathcal S_{ij}\) consists of the support of \(G_{ij}\)
together with zero, which is an admissible type and report even if it
has probability zero. Write
\(\mathcal S_i=\prod_j\mathcal S_{ij}\) and
\(\mathcal S=\prod_i\mathcal S_i\) for the type spaces, and
\(G_i=\bigotimes_jG_{ij}\) and \(G_{i,-j}=\bigotimes_{k\ne j}G_{ik}\)
for the laws of \(s_i\) and \(s_{i,-j}\).

The public function \(f_i:\mathcal S_i\to\R_+\) determines buyer \(i\)'s
value for a bundle \(A\subseteq[m]\):
\[
 v_i(s_i,A)=f_i(s_{i,A},0_{-A}).
\]
Thus an item contributes to value only when it is included in the bundle.
We assume that \(f_i\) is coordinatewise nondecreasing, \(f_i(0)=0\), and
\emph{submodular}:
\[
 f_i(s_i)+f_i(r_i)\ge f_i(s_i\wedge r_i)+f_i(s_i\vee r_i)
 \qquad(s_i,r_i\in\mathcal S_i),
\]
where \(\wedge\) and \(\vee\) take coordinatewise minima and maxima.
Increasing the signal of one item therefore raises value by a weakly
smaller amount when the other signals are higher. In particular, the value of a bundle is
monotone and submodular for each type.

We measure each signal in units of the item's singleton value, so
\(f_i(s_{ij}e_j)=s_{ij}\), where \(e_j\) is the \(j\)-th unit
vector.\footnote{This normalization is available in a model with finite
raw signals and a monotone submodular value function. If the raw
signal of item \(j\) rises from \(a\) to \(b\), submodularity
bounds its effect on any bundle by \(f_i(be_j)-f_i(ae_j)\).
Signals with the same singleton value thus give the same bundle values
at every background. Merging them and using the singleton value as the
coordinate preserves independence and submodularity.}
For example, \(f_i(s_i)=\sum_js_{ij}\) gives \emph{additive} values, and
\(f_i(s_i)=\max_js_{ij}\) gives \emph{unit-demand} values. Buyers evaluate
randomized allocations by expected value.

\subsection{The benchmark}
\label{sec:mechanisms}

By the revelation principle \citep{Myerson81}, we can compare mechanisms
through their truthful direct implementations. A randomized direct
mechanism \(M\) specifies a lottery over bundle assignments and
\emph{nonnegative payments} \(p_i(r)\ge0\) at every report profile
\(r\in\mathcal S\). Ex-post feasibility requires the assigned bundles
\(S_1(r),\ldots,S_n(r)\) to be pairwise disjoint in every realization.

When buyer \(i\) has type \(s_i\) and reports \(r_i\), her interim gross
value, allocation probabilities, and payment are
\[
 \begin{aligned}
 w_i(s_i;r_i)&=\E\bigl[v_i(s_i,S_i(r_i,s_{-i}))\bigr],\\
 x_{ij}(r_i)&=\Prb\bigl(j\in S_i(r_i,s_{-i})\bigr),\\
 p_i(r_i)&=\E_{s_{-i}}\bigl[p_i(r_i,s_{-i})\bigr].
 \end{aligned}
\]
The first two expressions average over opponents' types and the
mechanism's lottery. The third uses the same symbol for the interim payment
as for the payment at a report profile; the argument indicates which is
meant. Write \(w_i(s_i)=w_i(s_i;s_i)\) for truthful gross
value and \(u_i(s_i)=w_i(s_i)-p_i(s_i)\) for interim utility. We require
the following conditions at all types, including those of probability
zero:\footnote{This is without loss. A mechanism that satisfies the
conditions on the support extends to the other types by giving each her
preferred supported report, or nothing at zero payment.}
\begin{enumerate}[label=(\alph*),leftmargin=2.4em,itemsep=1pt]
 \item \emph{Bayesian incentive compatibility} (BIC). For every buyer
 \(i\) and every \(s_i,r_i\in\mathcal S_i\),
 \[
  u_i(s_i)\ge w_i(s_i;r_i)-p_i(r_i).
 \]
 \item \emph{Interim individual rationality}. For every \(i,s_i\),
 \(u_i(s_i)\ge0\).
\end{enumerate}
Throughout, \(M\) denotes a feasible mechanism satisfying these conditions
and the restriction on payments; we call any such \(M\) a
\emph{benchmark mechanism}. Its expected consumer surplus is
\[
 \CS(M)=\sum_i\E[u_i(s_i)],
 \qquad
 \OPT=\max_M\CS(M).
\]
Nonnegative payments and participation give
\(0\le u_i(s_i)\le w_i(s_i)\le v_i(s_i,[m])\), so the objective is
bounded. The maximum exists because types and feasible assignments are
finite.\footnote{For this existence argument, replace each payment
\(p_i(r_i,r_{-i})\) by its interim expectation \(p_i(r_i)\). This preserves
incentives, interim participation, and consumer surplus, and payments
remain nonnegative at every profile. Participation bounds these payments
by the largest full-bundle value. The resulting allocation and payment
variables lie in a compact set defined by finitely many linear constraints.}
A mechanism, or the better of several mechanisms, provides an
\emph{\(\alpha\)-approximation} if its expected consumer surplus is at
least \(\OPT/\alpha\).

\subsection{The two mechanisms}
\label{sec:two-mechanisms}

Separate selling uses burnt payments to screen buyers item by item.
Artificial currency instead gives buyers budgets that they can direct
toward the items they value most, without making real payments.

\subsubsection{Selling separately}
\label{sec:selling-separately}

For item \(j\), buyer \(i\)'s singleton value is \(s_{ij}\sim G_{ij}\).
An \emph{item auction} is a deterministic single-item mechanism with
nonnegative payments that is dominant-strategy incentive compatible and
ex-post individually rational. Thus, holding rivals' reports fixed,
truthful reporting maximizes the buyer's utility and gives her
nonnegative utility. Write \(a_{ij}(t)\in\{0,1\}\) and \(p_{ij}(t)\ge0\)
for the allocation and payment at reported singleton values
\(t\in\prod_i\mathcal S_{ij}\). Each auction awards the item to at most
one buyer.

Let \(\MB_j\) be the maximum expected consumer surplus of an item
auction for \(j\). These stronger incentive and participation conditions
do not reduce the single-item Bayesian optimum \citep{HR08}.\footnote{For
finite and possibly asymmetric priors, this also follows from the
implementation step in Appendix~\ref{app:scalar}.} Let \(\SEP\) denote the sum of these single-item optima:
\[
 \SEP=\sum_j\MB_j.
\]
In \emph{unrestricted separate selling}, we run one auction per item.
Each buyer receives all the items she wins and pays their total price.
With additive values, truthful reporting maximizes her utility in each
auction and across items, so the optimal expected consumer surplus is
\(\SEP\), and \(\SEP\le\OPT\).

For general values, we use \emph{one-sale separate selling}: separate selling with at most one sale
per buyer. Run one item auction per item and treat each award as a
tentative offer. Among buyer \(i\)'s offers, give her the item with the
largest strictly positive reported surplus
\(t_{ij}-p_{ij}(t_{\cdot j})\), breaking ties in a fixed order. She pays
only for that item. A buyer with no profitable offer receives nothing and
pays zero. Canceled awards leave the items unassigned. We call this
selection the \emph{best-offer rule}. Write \(\SEP_1\)
for the maximum expected consumer surplus over these mechanisms.
The item auctions are chosen jointly for this best-offer objective;
they need not maximize their individual surpluses. Both maxima exist:
there are finitely many allocation rules, and ex-post participation
bounds payments at every profile.

\begin{lemma}[One-sale separate selling is truthful]
\label{lem:one-sale}
For every collection of item auctions, one-sale separate selling is
dominant-strategy incentive compatible and ex-post individually rational.
Consequently, \(\SEP_1\le\OPT\).
\end{lemma}

Appendix~\ref{app:one-sale} gives the proof.
The restriction to one sale concerns the mechanism: buyers may
value bundles, and the benchmark may allocate several items to a buyer.

\subsubsection{Artificial currency}

\begin{definition}[Artificial-currency mechanism]
\label{def:artificial-currency}
An \emph{artificial-currency mechanism} gives each buyer \(i\) a personalized
budget \(B_i\ge0\) of \emph{credits}, a currency with no value outside the mechanism.
Budgets are fixed before signals are realized. After observing \(s_i\),
buyer \(i\) distributes her credits across items by choosing
\[
 b_i\in\mathcal A_i=\Bigl\{b_i\in\R_+^m:\ \textstyle\sum_jb_{ij}\le B_i\Bigr\},
\]
and an allocation rule maps each credit profile \(b=(b_1,\ldots,b_n)\) to
a lottery over feasible assignments. Credits cannot be exchanged, saved, or transferred, and the
mechanism collects no monetary payments.
\end{definition}

Two properties of this class matter below: a buyer's feasible actions do
not depend on her type, and nothing is paid. We use the following rule.

\begin{definition}[Proportional credit mechanism]
\label{def:credit-game}
The \emph{proportional credit mechanism} \(\Gamma(B,h)\) is the
artificial-currency mechanism with budgets
\(B_i\ge0\), not all zero, and item seeds \(h_j>0\). Each buyer receives
a fixed seed credit \(h_j/n\) on item \(j\), which does not count against
her budget.
At a credit profile \(b\), each item \(j\) is awarded to exactly one
buyer, with probabilities
\begin{equation}
 \chi_{ij}(b)=\frac{h_j/n+b_{ij}}{h_j+\sum_kb_{kj}}.
 \label{eq:credit-allocation}
\end{equation}
Conditional on \(b\), allocations are independent across items. All
payments are zero. Write \(\barB=\sum_iB_i\) and \(\barh=\sum_jh_j\).
\end{definition}

A buyer's winning probability in \eqref{eq:credit-allocation} equals her
share of the credits on the item. Positive seeds make payoffs continuous
in bids and ensure equilibrium existence. Their total mass can be arbitrarily small relative to the
total budget. Multiplying all budgets, seeds, and bids by the same positive
factor preserves allocations and equilibria; only their relative sizes
matter.

Since payments are zero, buyer \(i\)'s expected payoff at credit profile
\(b\) is \(W_i(s_i,b)=\E[v_i(s_i,S_i)\mid b]\). A strategy
\(\sigma_i\) assigns a probability distribution on \(\mathcal A_i\) to each type
\(s_i\). Opponents' types and bids induce a distribution of opposing bids
that does not depend on \(s_i\), by independence across buyers. A profile
\(\sigma\) is a \emph{Bayes--Nash equilibrium} if each type's mixed action
maximizes her expected payoff:
\begin{equation}
 g_i(s_i):=\E_{\sigma}[W_i(s_i,b)\mid s_i]
 =\max_{b_i\in\mathcal A_i}\E_{\sigma_{-i}}[W_i(s_i,(b_i,b_{-i}))].
 \label{eq:bne}
\end{equation}
We choose best responses also at types of probability zero, whose bids
do not affect the distribution of play. Write \(\BNE(B,h)\) for the set
of equilibria and suppress the dependence of \(g_i\) on \(\sigma\).

\begin{lemma}[Equilibrium existence]
\label{lem:equilibrium-existence}
Every proportional credit mechanism \(\Gamma(B,h)\) has a Bayes--Nash
equilibrium.
\end{lemma}

Appendix~\ref{app:existence} gives the proof. Every equilibrium has a BIC
direct implementation with zero payments and the same consumer surplus,
\[
 \CS(B,h;\sigma)=\sum_i\E[g_i(s_i)],
\]
so \(\CS(B,h;\sigma)\le\OPT\).
Our guarantee will hold at every equilibrium of the chosen mechanism.

\section{The Approximation Guarantee}
\label{sec:main}

Separate selling and the proportional credit mechanism can each perform
poorly on some instances, but the better of the two always provides a
constant approximation to the unrestricted benchmark.

\subsection{Main results}
\label{sec:results}

\begin{theorem}[Submodular values]
\label{thm:submodular}
For every \(0<\eta\le10^{-3}\), there is a proportional credit mechanism
\(\Gamma(B,h)\) with \(\barh=\eta\barB\) such that
\[
 \max\{\SEP_1,\CS(B,h;\sigma)\}\ge\frac{\OPT}{3.80}
 \qquad\text{at every }\sigma\in\BNE(B,h).
\]
\end{theorem}

\begin{theorem}[Additive values]
\label{thm:main}
Suppose values are additive and let \(\eta>0\). There is a proportional
credit mechanism \(\Gamma(B,h)\) with \(\barh=\eta\barB\) such that
\[
 \max\{\SEP,\CS(B,h;\sigma)\}\ge\frac{\OPT}{3+\eta}
 \qquad\text{at every }\sigma\in\BNE(B,h).
\]
\end{theorem}

In both theorems, when \(\OPT>0\) the budgets are \(B_i=\widehat K_i\), the stable values
of an optimal benchmark mechanism defined in \cref{sec:decomposition}, and
\(\eta=\barh/\barB\) is the \emph{relative seed mass}; the seeds \(h_j\)
may be any positive numbers with this total.
Budgets and seeds are thus chosen from the prior before signals are realized.
For additive values, taking \(\eta\) arbitrarily small gives an
approximation factor arbitrarily close to \(3\). This limiting factor
need not be achieved by a single mechanism with positive seeds.

The additive guarantee is stronger because a buyer can realize the gains
from all the items she wins. With submodular values, winning one item can
reduce the value of another, so truthful bidding in each item auction
need not maximize utility across auctions. Letting each buyer take only
her best offer restores incentive compatibility (\cref{lem:one-sale}). It also replaces the
sum of her tentative surpluses by their maximum, so \(\SEP_1\le\SEP\).
The first theorem shows that this restricted format, combined with
credits, still gives a constant approximation.

\begin{remark}[Symmetric buyers and equal budgets]
\label{rem:symmetric}
Suppose buyers are ex-ante symmetric: the value function \(f_i\) and the
signal distributions \((G_{ij})_{j\in[m]}\) do not depend on \(i\).
Relabeling the buyers of an optimal benchmark mechanism gives another
optimal benchmark mechanism, and so does the uniform mixture over all
relabelings, because mixtures preserve feasibility, incentive
compatibility, individual rationality, and consumer surplus. In this
symmetric mechanism every buyer has the same stable value. The budgets in
\cref{thm:submodular,thm:main} can therefore be taken equal across buyers,
as in the course-bidding systems discussed in the introduction. Since only
relative sizes matter, the credit mechanism is then prior-free: equal
budgets and arbitrary positive seeds of relative mass \(\eta\). The prior
enters only through the comparison with separate selling.
\end{remark}

\subsection{Neither mechanism suffices alone}
\label{sec:both-needed}

The need for both mechanisms already arises with additive values. The
first example gives buyers many opportunities to find an unusually
valuable item, while each item auction generates only bounded surplus.
The mechanism that performs well in this example uses no payments at all.

\begin{example}[Separate selling has no constant guarantee]
\label{ex:sep-fails}
Let \(n=m=N=2^L\), where \(L\ge1\) is an integer. Values are additive,
and all item values are independent draws of \(V\in\{0,\ldots,L\}\)
with \(\Prb(V\ge k)=2^{-k}\) for \(k=1,\ldots,L\). Then
\[
 \SEP_1\le\SEP\le2N,
 \qquad
 \OPT\ge N\beta_N^2L,
 \qquad \beta_N=1-(1-1/N)^N\ge1-1/e.
\]
Consequently, \(\SEP/\OPT\le2/(\beta_N^2L)\to0\) as \(L\to\infty\).
\end{example}

\begin{proof}
\emph{The single-item bound.} In any admissible single-item mechanism,
write \(u_i(k)\) and \(x_i(k)\) for buyer \(i\)'s interim utility and
allocation at value \(k\). Participation and nonnegative payments give
\(u_i(0)=0\). Letting type \(k-1\) imitate type \(k\) gives
\(u_i(k)-u_i(k-1)\le x_i(k)\). Since
\(\Prb(V\ge k)\le2\Prb(V=k)\),
\[
 \E u_i(V)\le\sum_{k=1}^L\Prb(V\ge k)x_i(k)
 \le2\E x_i(V).
\]
Summing over buyers and using feasibility bounds the item's surplus by
\(2\). Hence \(\SEP\le2N\).

\emph{A better multi-item mechanism.} Each buyer claims an item of
highest reported value, choosing uniformly among ties. Allocate each
claimed item uniformly among its claimants, with no payments, and leave
unclaimed items unallocated. Opponents' truthful claims are independent
and uniform across items, so every item offers the same winning
probability. Claiming a favorite item is therefore optimal.

An item is claimed with probability \(\beta_N\), so the expected number
of allocations is \(N\beta_N\). By symmetry, every buyer wins with
probability \(\beta_N\), regardless of her type and which item she
claims. Her maximum value equals \(L\) with probability \(\beta_N\),
because \(\Prb(V=L)=1/N\). Expected consumer surplus is therefore
\[
 N\beta_N\E[\max_{j\le N}s_{1j}]\ge N\beta_N^2L.
\]
This proves the benchmark lower bound and the vanishing ratio.
\end{proof}

With a single item and no payments, all strictly positive types rank
their choices solely by the probability of winning. Artificial currency
therefore cannot make a buyer more likely to win when her value is higher.
The argument uses only the two properties noted after
\cref{def:artificial-currency}, so the bound in the next example applies
to every mechanism without payments.

\begin{example}[Artificial currency has no constant guarantee]
\label{ex:credit-fails}
Let \(m=1\) and \(n=N\ge2\). Buyers have i.i.d.\ values
\[
 s_{i1}=\begin{cases}
 N&\text{with probability }1/N,\\
 1&\text{with probability }1-1/N.
 \end{cases}
\]
Every equilibrium of any artificial-currency mechanism has expected
consumer surplus at most \(2-1/N\), while
\(\OPT\ge(N-1)\beta_N\), where \(\beta_N=1-(1-1/N)^N\ge1-1/e\).
The ratio tends to zero as \(N\to\infty\), uniformly over such
mechanisms and their equilibria.
\end{example}

\begin{proof}
Fix any artificial-currency mechanism and one of its equilibria.
Independence makes opponents' play independent of buyer \(i\)'s value.
With no payments and type-independent feasible actions, types \(1\)
and \(N\) maximize the same winning probability. Each can imitate the
other's strategy, so their interim winning probabilities coincide.
Denote this probability by \(\chi_i\). Feasibility gives
\(\sum_i\chi_i\le1\), and expected consumer surplus is at most
\[
 \sum_i\chi_i\E[s_{i1}]
 =\Bigl(2-\frac1N\Bigr)\sum_i\chi_i\le2-\frac1N.
\]

Money burning can screen out low values. Allocate the item to the first
buyer in a fixed order who reports \(N\), charging that buyer \(1\)
and charging all others zero. This direct mechanism is dominant-strategy
incentive compatible and ex-post individually rational, including at
the zero type. A buyer with value \(N\) exists with probability
\(\beta_N\), giving expected consumer surplus \((N-1)\beta_N\).
\end{proof}

\subsection{Proof architecture: stable part and variation}
\label{sec:decomposition}

Fix a benchmark mechanism \(M\), and let \(s_i'\) be an independent copy
of buyer \(i\)'s type. Her utility has a stable part and a variation:
\[
 K_i=\E\min\{u_i(s_i),u_i(s_i')\},
 \qquad
 D_i=\tfrac12\E|u_i(s_i)-u_i(s_i')|.
\]
These quantities, and those defined next, depend on \(M\); we suppress
this dependence.
The identity \(\min\{a,b\}=(a+b-|a-b|)/2\) gives
\(\E u_i=K_i+D_i\). Thus the stable part is expected utility less its
dispersion across two independent types.\footnote{In concurrent work,
\citet{GLT26} derive the same decomposition for matroid-constrained
additive buyers.}

For the credit comparison, we replace the stable part by an upper bound
computed from gross value, which we call the buyer's \emph{stable value}:
\[
 \widehat K_i=\bigl(\E\sqrt{w_i(s_i)}\bigr)^2,
 \qquad \widehat K=\sum_i\widehat K_i,
 \qquad D=\sum_iD_i.
\]
Indeed, \(\min\{a,b\}\le\sqrt{ab}\), independence, and nonnegative payments imply
\[
 K_i\le\E\sqrt{u_i(s_i)u_i(s_i')}
 =\bigl(\E\sqrt{u_i(s_i)}\bigr)^2\le\widehat K_i.
\]
By Jensen's inequality, \(\widehat K_i\le\E[w_i(s_i)]\), and the gap is
large when gross value is concentrated on unlikely types. Summing over buyers gives
\begin{equation}
 \CS(M)\le\widehat K+D.
 \label{eq:surplus-decomposition}
\end{equation}

The two examples of \cref{sec:both-needed} sit at opposite ends of this
decomposition. In \cref{ex:sep-fails}, the payment-free mechanism in the
proof gives each buyer interim utility \(\beta_N\max_js_{ij}\), which
equals \(\beta_NL\) with probability \(\beta_N\). Her stable value is
therefore at least \(\beta_N^3L\), a fraction \(\beta_N^2\) or more of her
expected utility: credits can do well, while separate selling does not. In
\cref{ex:credit-fails}, the money-burning mechanism in the proof gives
buyer \(i\) gross value \(N(1-1/N)^{i-1}\) when her value is \(N\), an
event of probability \(1/N\), and zero otherwise. Hence
\(\widehat K_i=(1-1/N)^{i-1}/N\) and \(\widehat K=\beta_N\), while the
mechanism's consumer surplus is \((N-1)\beta_N\): almost all of its surplus
is variation, which credits cannot capture and separate selling can.

Credits capture stable value because a fixed budget gives the buyer
purchasing power across all her types. Giving each buyer a budget
proportional to \(\widehat K_i\) yields the first bound.

\begin{lemma}[Credits capture stable value]
\label{lem:credit-core}
Fix a benchmark \(M\) with \(\widehat K>0\) and let \(\eta>0\). Set
\(B_i=\widehat K_i\) and choose any positive seeds satisfying
\(\barh=\eta\widehat K\). Then
\[
 \CS(B,h;\sigma)\ge\frac{\widehat K}{2+\eta}
 \qquad\text{at every }\sigma\in\BNE(B,h).
\]
\end{lemma}

Separate selling helps when a buyer's utility varies with her signals.
With additive values, its surplus covers the entire variation. For
submodular values, one-sale separate selling covers the variation up to an
additional term controlled by stable value.

\begin{lemma}[Separate selling and the variation]
\label{lem:variation-charge}
For every benchmark mechanism,
\[
 D\le0.382\,\widehat K+1.03\,\SEP_1.
\]
If values are additive, then \(D\le\SEP\).
\end{lemma}

The constants \(0.382\) and \(1.03\) come from
\cref{lem:independent-selection}.

\Cref{sec:deviation} proves the credit bound, and \cref{sec:variation}
proves the variation bound. Together they give both theorems.

\begin{proof}[Proof of \cref{thm:submodular,thm:main}]
If \(\OPT=0\), any proportional credit mechanism with the required
seed mass suffices. Otherwise, choose an optimal benchmark \(M\).
Since its expected utility is positive, \(\widehat K>0\). Use the
budgets and seeds in \cref{lem:credit-core}, and fix any equilibrium
\(\sigma\). By \eqref{eq:surplus-decomposition} and the two lemmas,
\[
 \begin{aligned}
 \OPT
 &\le1.382\,\widehat K+1.03\,\SEP_1\\
 &\le1.382(2+\eta)\CS(B,h;\sigma)+1.03\,\SEP_1\\
 &\le(3.794+1.382\eta)\max\{\SEP_1,\CS(B,h;\sigma)\}.
 \end{aligned}
\]
The coefficient is below \(3.80\) for \(0<\eta\le10^{-3}\).
For additive values, the stronger variation bound instead gives
\[
 \OPT\le\widehat K+\SEP
 \le(2+\eta)\CS(B,h;\sigma)+\SEP
 \le(3+\eta)\max\{\SEP,\CS(B,h;\sigma)\}.
\]
Both inequalities hold at every equilibrium of the chosen mechanism.
\end{proof}

\section{Artificial Currency and Stable Value}
\label{sec:deviation}

A buyer can use her credits to pursue a bundle from the benchmark
allocation. We first bound the value she can secure at the prices
implied by opposing bids, then use feasibility to bound the total cost
of the benchmark bundles. This yields budgets that capture stable
value at every equilibrium.

\subsection{Credits as a market}

We view a buyer's credit budget as purchasing power over allocation
probabilities. Opponents' equilibrium spending determines the competition
she faces for each item, which we summarize through personalized implicit
prices.

Fix a proportional credit mechanism \(\Gamma(B,h)\) and an equilibrium
\(\sigma\in\BNE(B,h)\) with interim utilities \(g_i\). Let \(\bar b_{kj}\) be
buyer \(k\)'s expected spending on item \(j\), and define buyer \(i\)'s
implicit price \(q_{ij}\) by
\[
 \bar b_{kj}=\E_{\sigma}[b_{kj}],
 \qquad
 q_{ij}=h_j+\sum_{k\ne i}\bar b_{kj}>0 .
\]
The price includes the seed mass and other buyers' expected spending,
excluding buyer \(i\)'s own spending. Independence across buyers makes
opposing spending independent of \(s_i\), so she faces the same prices
at every type.

A buyer can use her budget to pursue any target bundle at these prices.
The \emph{implicit cost} of a bundle \(A\) is \(c_i(A)=\sum_{j\in A}q_{ij}\).

\begin{lemma}[Bundle deviation]
\label{lem:bundle-deviation}
Let \(\sigma\in\BNE(B,h)\). For every buyer with \(B_i>0\), every type \(s_i\), and every nonempty target bundle \(A\),
\begin{equation}
 g_i(s_i)\ge\frac{B_i}{B_i+c_i(A)}\,v_i(s_i,A).
 \label{eq:bundle-bound}
\end{equation}
\end{lemma}

\begin{proof}
The buyer can spend her
budget on the bundle in proportion to its item prices by bidding
\[
 d_{ij}=\frac{B_iq_{ij}}{c_i(A)}\mathbf 1\{j\in A\}.
\]
For each \(j\in A\), discarding her own seed share and applying Jensen's
inequality to the convex function \(y\mapsto d_{ij}/(d_{ij}+y)\) gives
a winning probability of at least
\[
 \E_{\sigma_{-i}}\!\left[
  \frac{d_{ij}}{d_{ij}+h_j+\sum_{k\ne i}b_{kj}}\right]
 \ \ge\ \frac{d_{ij}}{d_{ij}+q_{ij}}
 =\frac{B_i}{B_i+c_i(A)} .
\]
The denominator is \(B_i+c_i(A)\) rather than \(c_i(A)\) because her bids raise both
her share of the credits and the total credits on each item.

To convert these probabilities into a value guarantee, order the items
of \(A\) and let \(\ell_j\) be item \(j\)'s marginal contribution when it is
added to its predecessors. The weights are nonnegative and sum to
\(v_i(s_i,A)\). For any set \(S\), add the items of \(S\cap A\) in the
same order. Diminishing returns implies that each marginal contribution
is at least \(\ell_j\), so \(v_i(s_i,S)\ge\sum_{j\in S\cap A}\ell_j\).
The expected payoff from the deviation is therefore at least
\(B_i v_i(s_i,A)/(B_i+c_i(A))\). This bound uses only the winning
probabilities of the target items, not the independence of allocations
across items. By \eqref{eq:bne}, equilibrium utility at
each type dominates the payoff from every feasible deviation.
\end{proof}

If the bundle costs no more than her budget, this deviation secures each
target item with probability at least one half, and hence at least half of
the bundle's value.

\begin{lemma}[Ratio smoothness]
\label{lem:ratio-smoothness}
Fix budgets \(B_i\ge0\) with \(\barB>0\), seeds \(h_j>0\) and an equilibrium
\(\sigma\in\BNE(B,h)\) with interim utilities \(g_i\). For any benchmark
mechanism with interim gross values \(w_i\), interpreting \(0/0\) as zero,
\[
 \sum_{i:\,B_i>0}B_i\,
 \E_{s_i}\!\left[\frac{w_i(s_i)}{g_i(s_i)}\right]\le 2\barB+\barh .
\]
For every positive-budget buyer, \(g_i(s_i)>0\) whenever \(w_i(s_i)>0\).
\end{lemma}

\begin{proof}
Fix a buyer with \(B_i>0\). Rearranging the bound in
\cref{lem:bundle-deviation}
gives
\[
 B_i v_i(s_i,A)\le g_i(s_i)\bigl(B_i+c_i(A)\bigr)
 \qquad\text{for every }A\subseteq[m].
\]
The inequality for the empty bundle holds because \(v_i(s_i,\emptyset)=0\).
Average over the bundle drawn for buyer \(i\) by the benchmark conditional
on \(s_i\). Since the cost is additive across items, this yields
\[
 B_i w_i(s_i)\le g_i(s_i)\left(B_i+\sum_jq_{ij}x_{ij}(s_i)\right).
\]
In particular, positive gross value implies positive equilibrium utility.
Divide by \(g_i(s_i)\), interpreting \(0/0\) as zero. Taking
expectations and summing over the positive-budget buyers gives
\[
 \sum_{i:\,B_i>0}B_i\E_{s_i}\!\left[\frac{w_i(s_i)}{g_i(s_i)}\right]
 \ \le\ \barB+\sum_{i,j}q_{ij}\bar x_{ij},
 \qquad \bar x_{ij}=\E_{s_i}[x_{ij}(s_i)].
\]
The last sum includes all buyers. Benchmark feasibility gives
\(\sum_i\bar x_{ij}\le1\), and \(\sum_j \bar b_{kj}\le B_k\) by the budget
constraint. Thus
\[
 \sum_{i,j}q_{ij}\bar x_{ij}
 =\sum_jh_j\sum_i\bar x_{ij}+\sum_{j,k}\bar b_{kj}\sum_{i\ne k}\bar x_{ij}
 \ \le\ \barh+\barB .
\]
\end{proof}

The constant \(2\barB+\barh\) has a simple origin. One \(\barB\) is the
buyers' own budgets, which appear in the denominator of
\eqref{eq:bundle-bound}. The remaining \(\barB+\barh\) bounds the total
implicit cost of the benchmark's bundles: each item is allocated at most
once, and the credits that price it are the rivals' spending and the seeds.

The argument extends the scaled-allocation deviation of
\citet[Theorem~1]{BGM22} to Bayesian bundle targets. It uses independence
across buyers to price a deviation and ex-ante feasibility to bound the
total cost of the targets.

\subsection{From a benchmark to a guarantee}

A fixed budget must serve the buyer across all her types. The ratio
bound therefore leads to the square-root average in \(\widehat K_i\),
rather than her mean gross value. Choosing budgets proportional to
these stable values lets us combine the buyers' guarantees. The proof
applies the Cauchy--Schwarz inequality twice: across the types of one
buyer, to pass from the ratios \(w_i/g_i\) to her stable value, and across
buyers, where budgets proportional to stable values match the weights in
\cref{lem:ratio-smoothness}.

\begin{proof}[Proof of \cref{lem:credit-core}]
Suppose \(\widehat K>0\) and set \(B_i=\widehat K_i\), so
\(\barB=\widehat K\). Fix positive seeds with \(\barh=\eta\widehat K\)
and any \(\sigma\in\BNE(B,h)\). Write \(I=\{i:\widehat K_i>0\}\) and, for
\(i\in I\), define \(R_i=\E_{s_i}[w_i(s_i)/g_i(s_i)]\) and
\(\bar g_i=\E_{s_i}[g_i(s_i)]\). By \cref{lem:ratio-smoothness},
\[
 \sum_{i\in I}\widehat K_iR_i\le(2+\eta)\widehat K.
\]
The same lemma gives \(g_i(s_i)>0\) wherever \(w_i(s_i)>0\),
so \(0<R_i<\infty\) for \(i\in I\). Cauchy--Schwarz gives
\[
 \bar g_iR_i
 \ \ge\ \bigl(\E_{s_i}\sqrt{w_i(s_i)}\bigr)^2=\widehat K_i.
\]
Here we write \(\sqrt{w_i}=\sqrt{g_i}\sqrt{w_i/g_i}\) with the same ratio
convention as above. Dividing by \(R_i\), summing, and applying weighted
Cauchy--Schwarz gives
\[
 \CS(B,h;\sigma)\ge\sum_{i\in I}\frac{\widehat K_i}{R_i}
 \ \ge\ \frac{\widehat K^2}{\sum_{i\in I}\widehat K_iR_i}
 \ \ge\ \frac{\widehat K}{2+\eta}.\qedhere
\]
\end{proof}

\section{Separate Selling and the Variation}
\label{sec:variation}

Separate selling screens buyers one item at a time. To prove the
variation bounds in \cref{lem:variation-charge}, we connect its surplus
to benchmark utility through the utility difference obtained by setting
one item's signal to zero.

Fix a benchmark mechanism \(M\). For each buyer \(i\) and item \(j\),
define the \emph{coordinate utility difference}
\[
 \delta_{ij}(s_i)=u_i(s_i)-u_i(0,s_{i,-j}).
\]
Incentive compatibility and monotonicity of values make \(u_i\)
coordinatewise nondecreasing: a higher type can imitate a lower type
and obtain at least the lower type's utility. Thus \(\delta_{ij}\ge0\).

The same incentive constraints also bound each utility difference by the
corresponding difference in gross value. If type \((0,s_{i,-j})\) reports
\(s_i\), she receives that type's allocation and pays its payment.
Incentive compatibility therefore gives
\begin{equation}
 \delta_{ij}(s_i)\le
 \E\bigl[v_i(s_i,S_i)-v_i((0,s_{i,-j}),S_i)\mid s_i\bigr],
 \label{eq:coordinate-loss-value}
\end{equation}
where \(S_i\) is the benchmark bundle. For any fixed bundle, this value
difference is zero unless it contains \(j\), and otherwise equals the
drop in value from removing \(j\).

Coordinate utility differences add up to at most gross value.
For any fixed bundle, submodularity bounds the drop in
value from removing an item by the value added by that item in any
ordering of the bundle. These added values sum to the bundle's value.
The bound on coordinate utility differences
\eqref{eq:coordinate-loss-value} therefore implies
\begin{equation}
 \sum_j\delta_{ij}(s_i)\le w_i(s_i).
 \label{eq:coordinate-loss-mass}
\end{equation}

\subsection{Additive values: summing the item surpluses}
\label{sec:additive-variation}

For additive values, we prove \(D\le\SEP\) by first bounding variation
by the sum of expected coordinate utility differences. We then show that
separate selling covers this sum, item by item.

\begin{lemma}[Variation and coordinate utility differences]
\label{lem:variation-sum}
For every benchmark mechanism,
\begin{equation}
 D\le\sum_{i,j}\E[\delta_{ij}(s_i)].
 \label{eq:variation-sum}
\end{equation}
\end{lemma}

\begin{proof}
Start with the two independent types
\(s_i,s_i'\sim G_i\) in the definition of \(D_i\).
Replace the coordinates of \(s_i\) by those of \(s_i'\)
one at a time. For two adjacent types \((t,r)\) and \((t',r)\) with
shared background \(r\), monotonicity gives
\[
 |u_i(t,r)-u_i(t',r)|
 \le [u_i(t,r)-u_i(0,r)]+[u_i(t',r)-u_i(0,r)].
\]
The background has its original product law and is independent of the
two draws of the coordinate being replaced. Taking expectations and applying
the triangle inequality along the replacement sequence, then summing
over buyers, yields \eqref{eq:variation-sum}.
\end{proof}

The proof uses only that \(u_i\) is coordinatewise nondecreasing and that
a buyer's item signals are independent. It does not use additivity.

\begin{lemma}[Item-by-item coverage]
\label{lem:item-coverage}
Suppose values are additive. For every benchmark mechanism and every item
\(j\),
\[
 \sum_i\E[\delta_{ij}(s_i)]\le\MB_j.
\]
\end{lemma}

\begin{proof}
We construct a feasible single-item mechanism whose
expected surplus equals the sum of the expected utility differences
associated with item \(j\). Write \((t,r)\) for a type with singleton
value \(t\) for \(j\) and other signals \(r\). Incentive compatibility
and additivity imply
\[
 u_i(t,r)\ge u_i(t',r)+(t-t')x_{ij}(t',r)
 \qquad(t,t'\in\mathcal S_{ij}).
\]
Define interim utility, allocation, and payment for the single-item
problem by
\[
 \begin{aligned}
 \widetilde u_{ij}(t)
 &=\E_{s_{i,-j}}[u_i(t,s_{i,-j})-u_i(0,s_{i,-j})],\\
 \widetilde x_{ij}(t)&=\E_{s_{i,-j}}[x_{ij}(t,s_{i,-j})],\\
 \widetilde p_{ij}(t)&=t\widetilde x_{ij}(t)-\widetilde u_{ij}(t).
 \end{aligned}
\]
Averaging the incentive inequality proves single-item incentive
compatibility. Monotonicity gives \(\widetilde u_{ij}\ge0\). With
additive values, the bound \eqref{eq:coordinate-loss-value} becomes
\(\delta_{ij}(t,r)\le t x_{ij}(t,r)\). Averaging this bound gives
\(\widetilde p_{ij}(t)\ge0\).

To implement these allocations, collect the scalar reports \(t_k\),
sample other signals \(\hat s_{k,-j}\sim G_{k,-j}\) independently for
each buyer, and run the benchmark on \((t_k,\hat s_{k,-j})_{k\in[n]}\).
Assign only item \(j\), charging buyer \(k\) the nonnegative payment
\(\widetilde p_{kj}(t_k)\) at every profile. Independence gives the
claimed interim allocations, and the benchmark supplies feasibility.
The single-item Bayesian optimum equals \(\MB_j\), so
\[
 \sum_i\E[\delta_{ij}(s_i)]
 =\sum_i\E[\widetilde u_{ij}(s_{ij})]\le\MB_j.\qedhere
\]
\end{proof}

Summing \cref{lem:item-coverage} over items and applying \cref{lem:variation-sum} proves the
additive part of \cref{lem:variation-charge}:
\[
 D\le\sum_{i,j}\E[\delta_{ij}(s_i)]\le\sum_j\MB_j=\SEP.
\]

\subsection{Submodular values: taking the best offer}
\label{sec:submodular-variation}

For submodular values, our goal is to prove
\(D\le0.382\,\widehat K+1.03\,\SEP_1\). One-sale separate selling gives
each buyer only the largest of her tentative surpluses, so we compare it with a
maximum of coordinate utility differences rather than their sum.

For each buyer, draw independent variables \(\delta_{ij}^{*}\), one
from the marginal law of each \(\delta_{ij}(s_i)\), and write
\[
 T_i=\E[\max_j\delta_{ij}^{*}],\qquad T=\sum_iT_i.
\]
The independent copies allow a comparison with offers from separate
item auctions even though the original utility differences may be
dependent. We first show that \(\SEP_1\) covers \(T\), and then bound each
buyer's variation \(D_i\) in terms of \(T_i\) and her stable value
\(\widehat K_i\).

\subsubsection{One-sale separate selling and the independent maximum}

\begin{lemma}[One-sale separate selling covers the independent maximum]
\label{lem:scalar-coverage}
For every benchmark mechanism, \(T\le\SEP_1\).
\end{lemma}

With one sale, the gain from an item depends on the other offers. If
their best surplus is \(z\), an item with surplus \(x\) adds
\((x-z)_+=\max\{x-z,0\}\). We therefore need a comparison above
every possible competing offer. The proof constructs lotteries over
feasible item auctions whose tentative surpluses \(Z_{ij}\) satisfy
\begin{equation}
 \E[(\delta_{ij}(s_i)-z)_+]\le\E[(Z_{ij}-z)_+]
 \qquad(z\ge0).
 \label{eq:offer-loss}
\end{equation}
That is, \(Z_{ij}\) dominates \(\delta_{ij}(s_i)\) in the increasing convex
order \citep{SS07}.
At \(z=0\), this is the comparison of means used for additive values;
positive thresholds compare the extra surplus available when the buyer
already has another offer.

The construction differs from the additive one in one respect. With additive values, averaging
allocation probabilities and utility differences over the other signals
already gives single-item incentive compatibility. With submodular values,
an item's allocation probability
can decrease as its signal rises, and averaging need not remove this
decrease. We restore monotonicity by \emph{ironing} before averaging. Holding the
other signals fixed, we pool adjacent signal values whenever their
allocation probabilities decrease, replacing the probabilities within
each pool by their prior-weighted mean. This makes allocation monotone
while preserving expected allocation within each pool. Preserving these
averages keeps the ironed allocations feasible: they are implemented by
sampling each buyer's other signals, redrawing her reported item signal
within its pool, and running the benchmark on the completed reports.

These feasible monotone interim allocations admit a lottery over
deterministic, dominant-strategy incentive-compatible item auctions.
Winners make the smallest nonnegative payment consistent with truthful
reporting, and losers pay zero. To establish \eqref{eq:offer-loss} for
the tentative surpluses \(Z_{ij}\) of these auctions, we use
submodularity to bound how quickly a coordinate utility difference can
grow. Combining this bound with the averaging within each pool gives the required surplus
comparison for every \(z\ge0\).

It remains to pass from single offers to best offers. We choose the auction lotteries independently
across items, so that each buyer's tentative surpluses
\(Z_{i1},\ldots,Z_{im}\) are independent, like the copies
\(\delta_{ij}^{*}\). Conditioning
on the other terms in the maximum, we can use
\eqref{eq:offer-loss} and the identity \(\max\{z,x\}=z+(x-z)_+\) to
replace each independent utility difference by its auction surplus inside
the maximum, one item at a time. This gives
\[
 T_i=\E[\max_j\delta_{ij}^{*}]\le\E[\max_jZ_{ij}].
\]
The right-hand side is buyer \(i\)'s expected surplus when the drawn
auctions are run with the best-offer rule, which gives her the largest of
her tentative surpluses. Summing over buyers therefore bounds \(T\) by the
expected consumer surplus of one-sale separate selling, averaged over the
auction lotteries.
Some fixed
collection of these deterministic item auctions achieves at least this
average surplus. It is
admissible for the maximization defining \(\SEP_1\), which proves the
lemma. Appendix~\ref{app:scalar} gives the details.

\subsubsection{Variation and the independent maximum}

The next lemma bounds \(D_i\) directly using \(T_i\) and stable
value \(\widehat K_i\), without relying on \eqref{eq:variation-sum}.

\begin{lemma}[Variation and the independent maximum]
\label{lem:variation-maximum}
For every benchmark mechanism and each buyer \(i\),
\[
 D_i\le0.382\,\widehat K_i+1.03\,T_i.
\]
\end{lemma}

The proof uses an intermediate statistic of the coordinate utility
differences.

Fix a vector \(z\in\R_+^m\). For each \(p\in(0,1)\), select each
entry independently with probability \(p\), and let \(Y_j(p)\) indicate
whether entry \(j\) is selected. The selected sum
\(\sum_jz_jY_j(p)\) has mean \(p\sum_jz_j\). Define
\[
 \mathcal B(z)=\int_0^1
 \E_Y\left|\sum_jz_j(Y_j(p)-p)\right|\,\frac{dp}{p}.
\]
Thus \(\mathcal B(z)\) integrates the expected absolute deviation of
the selected sum from its mean, with weight \(1/p\). The expectation
is only over the auxiliary Bernoulli variables.
A single entry contributes its own size, because
\(\E|Y_j(p)-p|=2p(1-p)\). Since \(\mathcal B\) is subadditive,
\(\mathcal B(z)\le\sum_jz_j\). When \(k\) entries are equal, however,
\(\mathcal B(z)\) grows only like \(\sqrt k\).

Applied to the vector of coordinate utility differences,
\(\delta_i(s_i)=(\delta_{ij}(s_i))_{j\in[m]}\), this statistic bounds a
buyer's variation.

\begin{lemma}[Product-space inequality]
\label{lem:product-space}
For every benchmark mechanism and each buyer \(i\),
\begin{equation}
 D_i\le\E_{s_i}\mathcal B(\delta_i(s_i)).
 \label{eq:variation-norm}
\end{equation}
\end{lemma}

The proof relies on an inequality for functions of independent coin
flips. Appendix~\ref{app:product-space} proves this inequality and applies
it using independence of the item signals and monotonicity of \(u_i\).
The comparison holds for any coordinatewise nondecreasing function on
our product type space; it does not use submodularity. Because
\(\mathcal B(z)\le\sum_jz_j\), \cref{lem:product-space} refines
\cref{lem:variation-sum}: the gain over the additive argument comes from
the square-root growth of \(\mathcal B\) when many coordinate utility
differences are of similar size.

\begin{lemma}[Independent selection and threshold counts]
\label{lem:independent-selection}
Let \(V=(V_1,\ldots,V_m)\) be a finitely supported random vector in
\(\R_+^m\), with arbitrary dependence among its coordinates. For
\(t\ge0\), define \(C_t(V)=\#\{j:V_j>t\}\). Then
\begin{equation}
 \E\mathcal B(V)
 \le0.382\int_0^\infty(\E\sqrt{C_t(V)})^2\,dt
      +1.03\int_0^\infty(1-e^{-\E C_t(V)})\,dt.
 \label{eq:variation-levels}
\end{equation}
\end{lemma}

Appendix~\ref{app:product-space} also proves \cref{lem:independent-selection}.
For a deterministic count, the value of \(\mathcal B\) at a vector of \(k\)
ones and \(m-k\) zeros is Ramanujan's \(Q\)-function, which grows like
\(\sqrt{\pi k/2}\) \citep{FGKP95}. The constants in
\eqref{eq:variation-levels} are derived from this growth rate.

\begin{proof}[Proof of \cref{lem:variation-maximum}]
For \(t\ge0\), let \(N_t(s_i)=\#\{j:\delta_{ij}(s_i)>t\}\) be the number
of coordinate utility differences that exceed \(t\). We prove that
\[
 \begin{aligned}
 D_i&\le\E\,\mathcal B(\delta_i(s_i))\\
 &\le0.382\int_0^\infty\bigl(\E\sqrt{N_t(s_i)}\bigr)^2\,dt
   +1.03\int_0^\infty\bigl(1-e^{-\E N_t(s_i)}\bigr)\,dt\\
 &\le0.382\,\widehat K_i+1.03\,T_i.
 \end{aligned}
\]
The first inequality is \cref{lem:product-space}. The second is
\cref{lem:independent-selection} applied to \(V=\delta_i(s_i)\), for
which \(C_t(V)=N_t(s_i)\). For the third, we bound the two integrals in
turn.

The first integral is controlled by gross value, which determines
\(\widehat K_i\).
Each utility difference is counted for thresholds between zero and its
size, so
\(\int_0^\infty N_t(s_i)\,dt=\sum_j\delta_{ij}(s_i)\le w_i(s_i)\)
by \eqref{eq:coordinate-loss-mass}.
Minkowski's integral inequality, applied to the functions
\(t\mapsto\sqrt{N_t(s_i)}\) in \(L^2(dt)\), states that the norm of their
average over \(s_i\) is at most the average of their norms. Squaring gives
\[
 \int_0^\infty(\E\sqrt{N_t(s_i)})^2\,dt
 \le\left(\E\sqrt{\sum_j\delta_{ij}(s_i)}\right)^2
 \le\left(\E\sqrt{w_i(s_i)}\right)^2
 =\widehat K_i,
\]
where the final equality is the definition of \(\widehat K_i\).

For the second integral, each copy \(\delta_{ij}^{*}\) exceeds \(t\) with
the same probability as \(\delta_{ij}(s_i)\), and these probabilities sum
to \(\E N_t(s_i)\). Since the copies are independent,
\[
 \Prb\bigl(\max_j\delta_{ij}^{*}>t\bigr)
 =1-\prod_j\bigl[1-\Prb(\delta_{ij}(s_i)>t)\bigr]
 \ge1-e^{-\E N_t(s_i)},
\]
where the inequality uses \(1-p\le e^{-p}\). Integrating this probability
over thresholds gives \(T_i\), which bounds the second integral.
\end{proof}

Summing over buyers and applying \cref{lem:scalar-coverage} yields
\[
 D\le0.382\,\widehat K+1.03\,T
 \le0.382\,\widehat K+1.03\,\SEP_1,
\]
completing the proof of \cref{lem:variation-charge}.

\section{Conclusions}
\label{sec:conclusion}

For buyers with monotone submodular values generated by independent
item signals with finite support, the better of one-sale separate selling and a proportional credit mechanism achieves a
\(3.80\)-approximation to optimal consumer surplus.
For additive values, letting buyers take all the items they win improves
the guarantee to a \((3+\eta)\)-approximation for every \(\eta>0\).
The factor approaches \(3\) as the relative seed mass \(\eta\) tends to zero, and the
comparison holds at every equilibrium of a proportional credit mechanism
fixed in advance.
The two devices capture complementary parts of benchmark surplus: artificial
currency covers the stable part, while burnt payments screen across value
levels and help control its variation.

A natural direction is to ask whether combining separate selling with
artificial currency remains effective beyond monotone submodular values and independent
distributions. With complements, the marginal value of an item
can increase as a buyer receives other items. Correlation within a buyer's
signals links the information revealed by screening across items; correlation
across buyers can make her own signals informative about the competition she
faces. Which valuation and information structures still permit a
constant-factor approximation to optimal consumer surplus using these two
devices?

\bibliographystyle{apalike}
\bibliography{refs}

\newpage
\appendix

\section{Separate Selling and Variation for Submodular Values}
\label{app:submodular}

\subsection{Truthfulness of one-sale separate selling}
\label{app:one-sale}

\begin{proof}[Proof of \cref{lem:one-sale}]
Fix a buyer and her rivals' reports. Losing reports pay zero. In an item auction with a winning report,
incentive compatibility makes all winning reports pay the same price. If the buyer's true singleton value exceeds this price, her
truthful report must win. Truthful reporting therefore makes every
strictly profitable offer available, and the best-offer rule gives her the
best one. A different report can only change which of these fixed-price
offers she receives and which one is chosen. It cannot give her more
utility. Her utility from truthful reporting is nonnegative because she
pays only for an offer with strictly positive surplus. Finally, a
dominant-strategy incentive-compatible and ex-post individually rational
mechanism with nonnegative payments is feasible for the benchmark, so
\(\SEP_1\le\OPT\).
\end{proof}

\subsection{One-sale separate selling and the independent maximum}
\label{app:scalar}

\begin{proof}[Proof of \cref{lem:scalar-coverage}]
We construct independent lotteries over item auctions whose tentative
surpluses \(Z_{ij}\) satisfy \eqref{eq:offer-loss}.
This comparison lets us replace each independent utility difference by
an offer inside the maximum. The proof has five steps.

\emph{Step 1: ironing the benchmark allocations.} Fix buyer \(i\), item \(j\), and
the buyer's other signals \(r\). Write
\(\phi_r(t)=u_i(t,r)-u_i(0,r)\) and \(x_r(t)=x_{ij}(t,r)\).
Submodularity bounds the effect of increasing the signal from
\(t'\) to \(t\) by \(t-t'\) on bundles containing \(j\), and by zero
on other bundles. Incentive compatibility and monotonicity therefore give
\[
 0\le \phi_r(t)-\phi_r(t')\le(t-t')x_r(t)
 \qquad(t\ge t').
\]

Let \(t_1<\cdots<t_q\) be the supported values of \(s_{ij}\), with
probabilities \(\pi_1,\ldots,\pi_q>0\). Write \(x_k=x_r(t_k)\),
\(\phi_k=\phi_r(t_k)\), and \(t_0=\phi_0=0\). The allocation probabilities
\(x_k\) need not be monotone. We iron them with the
pool-adjacent-violators procedure, starting with singleton
blocks and pooling adjacent blocks whenever their probability-weighted
means of \(x_k\) decrease. Write \(y_k=y_r(t_k)\) for the resulting
nondecreasing block means. Within every final block \([k_0,k_1]\), the
probability-weighted mean allocation is unchanged:
\[
 \sum_{k=k_0}^{k_1} \pi_ky_k=\sum_{k=k_0}^{k_1} \pi_kx_k.
\]

\emph{Step 2: feasibility of the ironed allocations.} For item \(j\), first consider reports in the support of each buyer's
prior. Independently draw each buyer's other signals from their prior and
redraw her reported item signal from its pooled block, using the prior
conditional on that block. Run the benchmark on the completed types and assign only
item \(j\). Under truthful reporting, the resampled types have their original
independent laws.
For a buyer with report \(t\) and sampled background \(r\), averaging
her allocation over the redraw within her block gives \(y_r(t)\).
Thus the interim allocations \(\bar y_i(t)=\E_r[y_r(t)]\) are jointly
feasible and nondecreasing.

\emph{Step 3: surplus from the ironed allocations.} Define
\[
 \Phi_r(t_k)=\Phi_k:=\sum_{\ell=1}^k(t_\ell-t_{\ell-1})y_\ell,
 \qquad \Phi_0=0.
\]

Every suffix of a pooled block \([k_0,k_1]\) has mean at most its block
mean \(\mu\): this holds initially and survives each merger, since a
suffix puts no more relative weight on the first, higher-mean block.
Thus \(\sum_{k=v}^{k_1} \pi_k(x_k-\mu)\le0\) for \(k_0\le v\le k_1\), and
in particular \(x_{k_1}\le \mu\). Suppose \(\phi_{k_0-1}\le \Phi_{k_0-1}\), which holds for the first block since
\(\phi_0=\Phi_0=0\). For \(k_0\le k\le k_1\), the incentive bound gives
\(\phi_k\le \phi_{k_0-1}+(t_k-t_{k_0-1})x_k\), while
\(\Phi_k=\Phi_{k_0-1}+(t_k-t_{k_0-1})\mu\). Subtracting yields
\[
 \phi_k-\Phi_k\le(t_k-t_{k_0-1})(x_k-\mu).
\]
At \(k=k_1\), the bound \(x_{k_1}\le \mu\) gives \(\phi_{k_1}\le \Phi_{k_1}\), completing
the induction.
For any \(z\ge0\),
\((\phi_k-z)_+-(\Phi_k-z)_+\le\mathbf 1_{\{\phi_k>z\}}(\phi_k-\Phi_k)\).
The factor \(\mathbf 1_{\{\phi_k>z\}}(t_k-t_{k_0-1})\) is nonnegative
and nondecreasing. Summation by parts using the suffix inequalities,
then summing over blocks, yields
\[
 \E_t[(\phi_r(t)-z)_+]\le\E_t[(\Phi_r(t)-z)_+].
\]
Introduce an auxiliary price \(C_r\), independent of \(t\), by putting mass
\(y_k-y_{k-1}\) at \(t_{k-1}\), with \(y_0=0\), and the remaining
mass at infinity. Its expected surplus at \(t_k\) is \(\Phi_k\).
Jensen's inequality therefore lets us replace \(\Phi_r(t)\) on the
right-hand side by the surplus \((t-C_r)_+\).

\emph{Step 4: dominant-strategy implementation.} We now implement the feasible
monotone interim allocations \(\bar y_i\) by item auctions. By
\citet[Theorem~5]{CDW12}, these allocations admit a finite mixture of
feasible deterministic rules, each nondecreasing in a buyer's own
report. Treat an unsupported zero report as the smallest supported
value. With rivals' reports fixed, charge every winning report
\(t_{k-1}\) if \(t_k\) is the buyer's lowest supported winning value;
charge losers zero. This is the smallest nonnegative winning payment
consistent with truthful reporting. These payments make every
deterministic rule dominant-strategy incentive compatible and ex-post
individually rational, so it is an item auction as defined in
\cref{sec:selling-separately}.

The winning price \(C_{ij}\), set to infinity if no report wins,
depends only on the rule and rivals' reports. It is independent of
the buyer's signal, and her tentative surplus is
\(Z_{ij}=(s_{ij}-C_{ij})_+\). The probability of first winning at \(t_k\) is the
increment of \(\bar y_i\) at that type, starting from zero. Thus
\(C_{ij}\) has the law of \(C_r\) mixed over backgrounds. Averaging
the preceding surplus comparison over backgrounds proves
\eqref{eq:offer-loss}.

The same implementation applies, without ironing, to the interim
allocations \(x_i\) of any Bayesian incentive-compatible and interim
individually rational single-item mechanism with nonnegative payments.
These allocations are nondecreasing, and the mechanism's interim utilities
satisfy \(u_i(t_k)\le\sum_{\ell\le k}(t_\ell-t_{\ell-1})x_i(t_\ell)\),
which is the expected surplus of type \(t_k\) in the resulting lottery
over item auctions. Some auction in the lottery therefore has at least the
mechanism's expected consumer surplus. This is the claim about \(\MB_j\)
in \cref{sec:selling-separately}.

\emph{Step 5: taking the best offer.} Draw the auction rules independently across
items. For each buyer, the \(Z_{ij}\) are independent because item
signals and rule draws are independent. Since
\(\max\{z,x\}=z+(x-z)_+\), conditioning on the other terms in the maximum
and applying \eqref{eq:offer-loss} replaces independent utility differences
by tentative surpluses one at a time:
\[
 T_i\le\E[\max_j Z_{ij}].
\]
The right-hand side is expected surplus under one-sale separate selling.
Summing over buyers bounds \(T\) by the mixture's expected surplus.
Some fixed collection of item auctions achieves at least this average
and is admissible for \(\SEP_1\). Hence \(T\le\SEP_1\).
\end{proof}

\subsection{Variation and the independent maximum}
\label{app:product-space}

\begin{proof}[Proof of \cref{lem:independent-selection}]
Set \(a=191/500\) and \(b=103/100\), and write \(c_0=\sqrt{\pi/2}\) and
\(d_0=c_0-1\).

The functional \(\mathcal B\) is homogeneous, subadditive, and bounded by
\(\sum_jz_j\). It is also coordinatewise nondecreasing: conditional
on the other Bernoulli variables, the expected absolute value is convex
in \(z_j\) and minimized at zero by Jensen's inequality.

Write
\(Q_k=\mathcal B(\mathbf1_A)\) for any \(A\subseteq[m]\) with
\(|A|=k\), where \(\mathbf1_A\) has \(k\) ones and \(m-k\) zeros.
The choice of \(A\) does not matter because the Bernoulli variables are
identically distributed, and \(Q_0=0\). The proof has four steps: a closed
form for \(Q_k\), the estimate \(Q_k\le c_0\sqrt k-d_0\), a bound for
random counts, and a layer-cake argument.

\emph{Step 1: a closed form.} For \(k\ge1\), we show that
\begin{equation}
 Q_k=\sum_{\ell=1}^k\frac{\ell!\binom{k}{\ell}}{k^\ell}
 =\int_0^\infty[1-e^{-k\psi(t)}]\,\frac{dt}{t^2},
 \qquad\psi(t)=t-\log(1+t).
 \label{eq:closed-form}
\end{equation}
The sum is Ramanujan's \(Q\)-function \citep{FGKP95}.
To see that \(Q_k\) equals this sum, let \(J_p\sim\operatorname{Bin}(k,p)\), and let
\([p^\ell]\) denote the coefficient of \(p^\ell\) in a polynomial.
Since \(\E(J_p-kp)=0\), the expected absolute deviation equals twice
the expected positive part. Expanding the binomial probabilities and
integrating gives
\[
 \begin{aligned}
 Q_k
 &=2\sum_{s=1}^k\binom{k}{s}\int_0^{s/k}
       (s-kp)p^{s-1}(1-p)^{k-s}\,dp\\
 &=\sum_{\ell=1}^k\frac{2[p^\ell]\E J_p^{\ell+1}}{\ell(\ell+1)k^\ell}
 =\sum_{\ell=1}^k\frac{\ell!\binom{k}{\ell}}{k^\ell}.
 \end{aligned}
\]
For the second equality in this display, expand \((1-p)^{k-s}\), integrate each power
using
\[
 \int_0^{s/k}(s-kp)p^{\ell-1}\,dp
 =\frac{s^{\ell+1}}{\ell(\ell+1)k^\ell},
\]
and group terms with the same power \(\ell\). The last equality counts
products using exactly \(\ell\) distinct Bernoulli variables, one twice:
their coefficient is \(\binom{k}{\ell}\ell!\binom{\ell+1}{2}\).
For the integral in \eqref{eq:closed-form},
expanding \((1+t)^{k-1}\) and integrating by parts gives
\[
 Q_k=k\int_0^\infty e^{-kt}(1+t)^{k-1}\,dt
 =\int_0^\infty[1-e^{-k\psi(t)}]\,\frac{dt}{t^2}.
\]

\emph{Step 2: the estimate \(Q_k\le c_0\sqrt{k}-d_0\) for \(k\ge1\).}
The inequality \(2\psi(t)\le t^2/(1+t)^{2/3}\) follows because the
difference is zero at zero and has derivative
\[
 \frac{2t}{1+t}\left[\frac{1+2t/3}{(1+t)^{2/3}}-1\right]\ge0
\]
by concavity of \(x^{2/3}\). Thus, with \(\tau=\sqrt{2\psi(t)}\),
we have \(\tau^3(1+t)/t^3\le1\). Since \(Q_1=1\), Tonelli's theorem
and this substitution give
\[
 Q_k-1
 =\frac12\int_1^k\int_0^\infty e^{-y\tau^2/2}
       \frac{\tau^3(1+t)}{t^3}\,d\tau\,dy
 \le\frac12\int_1^k\frac{c_0}{\sqrt y}\,dy
 =c_0(\sqrt k-1),
\]
as claimed.

\emph{Step 3: a bound for random counts.}
For a random count \(N\in\{0,\ldots,m\}\), we show that
\begin{equation}
 \E Q_N\le a(\E\sqrt N)^2+b(1-e^{-\E N}).
 \label{eq:random-count}
\end{equation}
Let \(\lambda=\E N>0\) and
\(\nu=\E\sqrt N\); the case \(\lambda=0\) is immediate. We have \(\nu\le \lambda\),
and Cauchy--Schwarz gives \(\nu^2\le \lambda\Prb(N>0)\). Thus, by Step~2,
\[
 \E Q_N-a\nu^2\le c_0\nu-(a+d_0/\lambda)\nu^2.
\]
If \(\lambda\le4/5\), this quadratic increases on \(0\le \nu\le \lambda\),
since \(c_0<63/50\) implies \(2a\lambda<2-c_0\). Hence it is at most \(\lambda-a\lambda^2\).
The fourth-order Taylor bound for \(e^{-\lambda}\) gives
\[
 b(1-e^{-\lambda})-\lambda+a\lambda^2\ge \lambda P(\lambda),\qquad
 P(\lambda)=\frac3{100}-\frac{133}{1000}\lambda
       +\frac{103}{600}\lambda^2-\frac{103}{2400}\lambda^3.
\]
On \([0,4/5]\), \(P''\ge1/8\), \(P(1/2)>1/1000\), and
\(|P'(1/2)|<7/1000\). Minimizing the resulting quadratic Taylor
lower bound gives \(P(\lambda)\ge1/1000-4(7/1000)^2>0\).

If \(\lambda\ge4/5\), maximizing the quadratic over all \(\nu\ge0\) and
using \(ab>\pi/8=c_0^2/4\) gives
\[
 \E Q_N-a\nu^2\le\frac{c_0^2}{4(a+d_0/\lambda)}
 \le b\frac{a\lambda}{a\lambda+d_0}\le b(1-e^{-\lambda}).
\]
The last inequality follows from
\[
 \frac{e^\lambda-1}{\lambda}\ge1+\frac \lambda2+\frac{\lambda^2}6+\frac{\lambda^3}{24}
 \ge\frac{191}{125}>\frac a{d_0},
\]
where \(d_0>1/4\). This proves \eqref{eq:random-count}.

\begin{samepage}
\emph{Step 4: a layer-cake argument.} By the definition of \(C_t(V)\),
\[
 V=\int_0^\infty(\mathbf1\{V_j>t\})_{j\in[m]}\,dt.
\]
Subadditivity and \eqref{eq:random-count} imply
\[
 \E\mathcal B(V)
 \le\int_0^\infty\E Q_{C_t(V)}\,dt
 \le a\int_0^\infty(\E\sqrt{C_t(V)})^2\,dt
      +b\int_0^\infty(1-e^{-\E C_t(V)})\,dt.
\]
\end{samepage}
\end{proof}

\begin{proof}[Proof of \cref{lem:product-space}]
Fix buyer \(i\), and write \(u=u_i\) and \(\delta_j=\delta_{ij}\).

We first establish an inequality for functions of independent signs,
by a semigroup interpolation in the spirit of \citet{IvHV20}.
Let \(\omega\) be uniform on \(\{-1,1\}^m\), and let
\(\omega^{(j)}\) denote its \(j\)-th sign flip. For
\(F:\{-1,1\}^m\to\R\), define
\(d_j^+F(\omega)=(F(\omega)-F(\omega^{(j)}))_+\) and
\(d^+F(\omega)=(d_j^+F(\omega))_{j\in[m]}\). We will show that
\[
 \tfrac12\E_\omega|F(\omega)-F(-\omega)|
 \le\E_\omega\mathcal B(d^+F(\omega)).
\]
To prove this, take any \(\gamma:\{-1,1\}^m\to[-1,1]\).
For \(0<\rho<1\), let \(\xi_j\) be independent signs of mean
\(\rho\), independent of \(\omega\), and put
\(P_\rho \gamma(\omega)=\E_\xi \gamma(\omega\xi)\), where
the product is coordinatewise. Differentiating the product probabilities
and pairing each sign vector with its flip gives
\[
 \begin{aligned}
 \frac{d}{d\rho}\E[F P_\rho \gamma]
 &=\frac1{2\rho}\E_\omega\sum_jd_j^+F(\omega)
       [P_\rho \gamma(\omega)-P_\rho \gamma(\omega^{(j)})]\\
 &=\frac{\E_{\omega,\xi}\bigl[\gamma(\omega\xi)
       \sum_jd_j^+F(\omega)(\xi_j-\rho)\bigr]}{1-\rho^2},
 \end{aligned}
\]
where the second equality follows by conditioning on \(\xi_j\).
Since \(P_0\gamma=\E \gamma\) and \(P_1\gamma=\gamma\), integration gives
\[
 \operatorname{Cov}(F,\gamma)
 \le\E_\omega\int_0^1\E_\xi
   \left|\sum_jd_j^+F(\omega)(\xi_j-\rho)\right|
   \frac{d\rho}{1-\rho^2}
 =\E_\omega\mathcal B(d^+F(\omega)).
\]
The last equality uses \(p=(1-\rho)/2\), \(\xi_j=1-2Y_j(p)\),
and symmetry under \(p\mapsto1-p\). Taking
\(\gamma(\omega)=\operatorname{sign}(F(\omega)-F(-\omega))\), with
\(\operatorname{sign}(0)=0\), gives
\(\E \gamma=0\) and makes this covariance equal to
\(\tfrac12\E_\omega|F(\omega)-F(-\omega)|\).

Now let \(s_i\) and \(s_i'\) be independent draws from the buyer's product
law \(G_i\). For each \(\omega\), let \(s^{\omega}\) be the type whose
\(j\)-th coordinate is that of \(s_i\) if \(\omega_j=1\) and that of
\(s_i'\) if \(\omega_j=-1\), and apply the preceding bound to
\(F(\omega)=u(s^{\omega})\), conditional on \(s_i\) and \(s_i'\).
Monotonicity gives \(d_j^+F(\omega)\le\delta_j(s^{\omega})\), since
setting that coordinate to zero can only lower utility. Coordinatewise
monotonicity of \(\mathcal B\) therefore gives
\(\mathcal B(d^+F(\omega))\le\mathcal B(\delta_i(s^{\omega}))\).
Unconditionally, \(s^{\omega}\) and \(s^{-\omega}\) are independent
draws from \(G_i\). Averaging over \(s_i\) and \(s_i'\) therefore proves
\eqref{eq:variation-norm}.
\end{proof}

\section{Equilibrium Existence}
\label{app:existence}

\subsection{Existence with positive seeds}

\begin{proof}[Proof of \cref{lem:equilibrium-existence}]
A pure strategy specifies one bid in the compact set \(\mathcal A_i\) for each
of buyer \(i\)'s finitely many types. Pure-strategy spaces are therefore
compact. Positive seeds make allocation probabilities continuous in
bids, and expected payoffs are finite sums of continuous functions.
Glicksberg's theorem \citep{Glicksberg52} gives a mixed-strategy Nash
equilibrium of this strategic-form game. Its marginal action
distributions at each type give behavioral strategies with the same
payoffs. Each positive-probability type must best respond, since
improving its action would improve ex-ante utility. At zero-probability
types, choose any best response, which exists by continuity and
compactness. These choices do not affect play at other types and give
the equilibrium in \eqref{eq:bne}.
\end{proof}

\subsection{Why seeds are needed}

Positive seeds ensure the payoff continuity used above. Their total mass
may be arbitrarily small relative to the total budget, but removing them
can destroy equilibrium existence even with deterministic additive values.
The following example is equivalent, up to relabeling buyers and a
positive rescaling of one buyer's values, to the additive example in
\citet{BGM22}. They rule out pure equilibria; the proof below also rules
out mixed equilibria.

\begin{samepage}
\begin{example}[Without seeds an equilibrium may fail to exist]
\label{ex:no-equilibrium}
Consider the unseeded rule obtained by setting \(h=0\) in
\eqref{eq:credit-allocation}, with the convention that an item receiving
no credits is left unallocated, so \(0/0:=0\) throughout this example.
Take \(n=m=2\), unit budgets \(B_1=B_2=1\), and deterministic additive
values with singleton-value vectors
\[
 s_1=(1,1)
 \qquad\text{and}\qquad
 s_2=(1,0).
\]
This mechanism has no Nash equilibrium, in pure or in mixed strategies.
\end{example}
\end{samepage}

\begin{proof}
Types are deterministic, so a strategy is a possibly random feasible
credit vector. Write \((b_{11},b_{12})\) for buyer \(1\)'s bid and
\((b_{21},b_{22})\) for buyer \(2\)'s bid.

Suppose first that \(\Prb(b_{11}>0)>0\). Buyer \(2\) only values
item \(1\), and her payoff \(\E[\theta/(\theta+b_{11})]\) from bidding
\(\theta\in[0,1]\) on it is strictly increasing: the integrand increases
whenever \(b_{11}>0\) and is nondecreasing otherwise. Her unique best
response is therefore \((b_{21},b_{22})=(1,0)\). Against this bid, buyer
\(1\)'s payoff from a feasible bid \((\alpha_1,\alpha_2)\) is
\[
 \frac{\alpha_1}{\alpha_1+1}+\mathbf1\{\alpha_2>0\}.
\]
Its supremum \(3/2\) is approached as \(\alpha_1\uparrow1\) and
\(\alpha_2\downarrow0\), but no feasible bid attains it: positive
\(\alpha_2\) forces \(\alpha_1<1\), while \(\alpha_2=0\) gives payoff
at most \(1/2\). Every mixed strategy also has expected payoff below
\(3/2\) and can be improved by a pure bid sufficiently close to the
supremum. Thus buyer \(1\) has no best response.

Suppose instead that \(b_{11}=0\) almost surely. Buyer \(1\) gets
nothing from item \(1\). Deviating to the feasible pure bid
\((1/2,1/2)\) gains at least \(1/3\) on that item, since \(b_{21}\le1\).
Her original winning probability on item \(2\) is at most
\(1/(1+b_{22})\), since \(b_{12}\le1\); after the deviation it is
\(1/(1+2b_{22})\). The decrease in her expected value from item \(2\)
is therefore at most
\[
 \frac{1}{1+b_{22}}-\frac{1}{1+2b_{22}}
 =\frac{b_{22}}{(1+b_{22})(1+2b_{22})}
 \ \le\ \frac{b_{22}}{(1+b_{22})^2}
 \ \le\ \frac14.
\]
The deviation improves her payoff by at least \(1/12\) for every
realization of the original bids, ruling out an equilibrium in this case
as well.
\end{proof}

\end{document}